\documentclass[11pt, a4paper,twocolumn]{article} %
\usepackage{siunitx}
\DeclareSIUnit\fps{fps}
\DeclareSIUnit\rpm{rpm}
\usepackage{microtype}

\usepackage[switch]{lineno} 

\usepackage{abstract}

\usepackage{geometry}
\newcommand{\f}{\mkern-2mu f\mkern-3mu}

\usepackage{xfp}
\usepackage{authblk}

\usepackage{latexsym}      
\usepackage{dcolumn}
\usepackage[hidelinks]{hyperref}
\hypersetup{
 colorlinks=false,
 citecolor=false,
 linkcolor=false,
 urlcolor=false,
 pdfpagemode=UseNone,
 pdfstartview=FitH}
\usepackage[utf8]{inputenc}
\usepackage{scalerel,stackengine,amsmath}

\usepackage[hidelinks]{hyperref}
\usepackage[citestyle=numeric,style=phys,biblabel=brackets,backend=biber,sorting=none, natbib]{biblatex}
\DefineBibliographyStrings{english}{andothers={\itshape et\addabbrvspace al\adddot}}

\usepackage[eulergreek]{sansmath}
\usepackage{upgreek}

\usepackage{filecontents}

\usepackage{booktabs}
\usepackage{dcolumn}
\newcolumntype{d}[1]{D{.}{.}{#1}}

\usepackage{bm}

\usepackage{xcolor}
\usepackage{dcolumn}

\usepackage{titlesec}
\titleformat{\section}
  {\bf\sffamily}
  {\thesection. }
  {5pt}
  {\MakeUppercase}
\renewcommand{\thesection}{\Roman{section}} 

\titleformat{\subsection}
  {\bf\sffamily}
  {\thesubsection. }
  {5pt}{}
  
\renewcommand{\thesubsection}{\Alph{subsection}}

\titleformat*{\paragraph}{\itshape}

\usepackage{caption}
\makeatletter
\providecommand*\AtBeginEnvironment[1]{%
  \@ifundefined{#1}%
    {\@latex@error{Environment #1 undefined}\@ehc
     \@gobble}%
    {\@ifundefined{ABE@env@#1}%
       {\expandafter\let\csname ABE@env@#1\expandafter\endcsname
          \csname #1\endcsname
        \expandafter\let\csname ABE@hook@#1\endcsname\@empty
        \@namedef{#1}{\@nameuse{ABE@hook@#1}\@nameuse{ABE@env@#1}}}%
       {}%
     \expandafter\g@addto@macro\csname ABE@hook@#1\endcsname}}
\@onlypreamble\AtBeginEnvironment
\makeatother

\usepackage{tikz}
\usepackage{pgfplots}
\pgfplotsset{compat=newest}

\usetikzlibrary{arrows}
\usepgfplotslibrary{fillbetween}
\usetikzlibrary{calc}
\usetikzlibrary{math}
\usetikzlibrary{patterns}
\usetikzlibrary{patterns.meta}
\usetikzlibrary{arrows.meta}
\usetikzlibrary{arrows}
\usetikzlibrary{positioning}
\usetikzlibrary{shapes, shapes.geometric, arrows}
\usetikzlibrary{arrows.meta}
\usetikzlibrary{patterns,patterns.meta}
\usetikzlibrary{intersections}
\usetikzlibrary{backgrounds}
\usepgfplotslibrary{colorbrewer}
\usepgfplotslibrary{statistics}
\pgfplotsset{
tick label style = {font=\sansmath\sffamily},
every axis label/.append style = {font=\sansmath\sffamily}
}
\def\centerarc[#1](#2)(#3:#4:#5)
{ \draw[#1] ($(#2)+({#5*cos(#3)},{#5*sin(#3)})$) arc (#3:#4:#5); }

\pgfplotscreateplotcyclelist{usualsuspectinner}{
{mark=*,c1},
{mark=triangle*,c4},
{mark=diamond*,gold},
{mark=halfcircle*,c11},
{mark=square*,c21}, 
{mark=halfsquare left*,c23},
{mark=pentagon*,c28},
      }
\pgfplotscreateplotcyclelist{usualsuspectouter}{
{mark=o,c1},
{mark=triangle,c4},
{mark=diamond,gold},
{mark=halfcircle,c11},
{mark=square,c21}, 
{mark=halfsquare left,c23},
{mark=pentagon,c28},
      }

\pgfplotsset{
    every first x axis line/.style={},
    every first y axis line/.style={},
    every first z axis line/.style={},
    every second x axis line/.style={},
    every second y axis line/.style={},
    every second z axis line/.style={},
    first x axis line style/.style={/pgfplots/every first x axis line/.append style={#1}},
    first y axis line style/.style={/pgfplots/every first y axis line/.append style={#1}},
    first z axis line style/.style={/pgfplots/every first z axis line/.append style={#1}},
    second x axis line style/.style={/pgfplots/every second x axis line/.append style={#1}},
    second y axis line style/.style={/pgfplots/every second y axis line/.append style={#1}},
    second z axis line style/.style={/pgfplots/every second z axis line/.append style={#1}}
}

\tikzset{
  font={\small\sffamily\sansmath}
}

\pgfplotsset{every tick label/.append style={font=\sffamily\sansmath}}
\pgfplotsset{every tick label/.append style={font=\small}}

\pgfplotsset{
    jja style/.style={
        boxplot/draw/whisker/.code 2 args={%
            \draw[/pgfplots/boxplot/every whisker/.try]
                (boxplot cs:##1) -- (boxplot cs:##2)
            ;
        },%
        boxplot/every box/.style={
            fill,
        },
        boxplot/draw direction=y,
        ylabel=\%,
        ylabel style={rotate=-90},
        boxplot/box extend=0.3,
    },
    rshift/.style={
        xshift=\pgfkeysvalueof{/pgfplots/rshift scale},
        legend image post style={xshift=-\pgfkeysvalueof{/pgfplots/rshift scale}},
        persianindigo,
    },
    lshift/.style={
        xshift=-\pgfkeysvalueof{/pgfplots/lshift scale},
        legend image post style={xshift=\pgfkeysvalueof{/pgfplots/lshift scale}},
        purple,
    },
    rshift scale/.initial=0.7em,
    lshift scale/.initial=0.7em,
}

\makeatletter
\def\pgfplots@drawaxis@outerlines@separate@onorientedsurf#1#2{%
    \if2\csname pgfplots@#1axislinesnum\endcsname
    \else
    \scope[/pgfplots/every outer #1 axis line,
        #1discont,decoration={pre length=\csname #1disstart\endcsname, post length=\csname #1disend\endcsname}]
        \pgfplots@ifaxisline@B@onorientedsurf@should@be@drawn{0}{%
            \draw [/pgfplots/every first #1 axis line] decorate {
                \pgfextra
                \pgfplotspointonorientedsurfaceabsetupfor{#2}{#1}{\pgfplotspointonorientedsurfaceN}%
                \pgfplots@drawgridlines@onorientedsurf@fromto{\csname pgfplots@#2min\endcsname}%
                \endpgfextra 
                };
        }{}%
        \pgfplots@ifaxisline@B@onorientedsurf@should@be@drawn{1}{%
            \draw [/pgfplots/every second #1 axis line] decorate {
                \pgfextra
                \pgfplotspointonorientedsurfaceabsetupfor{#2}{#1}{\pgfplotspointonorientedsurfaceN}%
                \pgfplots@drawgridlines@onorientedsurf@fromto{\csname pgfplots@#2max\endcsname}%
                \endpgfextra 
                };
        }{}%
    \endscope
    \fi
}%
\makeatother

\usepackage{sansmath}
\DeclareMathVersion{sans}
\SetSymbolFont{operators}{sans}{OT1}{cmbr}{m}{n}
\SetSymbolFont{letters}{sans}{OML}{cmbrm}{m}{it}
\SetSymbolFont{symbols}{sans}{OMS}{cmbrs}{m}{n}
\SetMathAlphabet{\mathit}{sans}{OT1}{cmbr}{m}{sl}
\SetMathAlphabet{\mathbf}{sans}{OT1}{cmbr}{bx}{n}
\SetMathAlphabet{\mathtt}{sans}{OT1}{cmtl}{m}{n}
\SetSymbolFont{largesymbols}{sans}{OMX}{iwona}{m}{n}

\usepackage{csvsimple}
\usepackage{changepage}
\usepackage[tbtags]{mathtools}

\usepackage{csquotes}
\usepackage{enumerate}
\usepackage{enumitem}
\usepackage{filecontents}

\usepackage[
	nonumberlist, 
	acronym, 
	nomain,
	nopostdot,
]{glossaries}

\usepackage{color, colortbl}

\usepackage{csvsimple}
\usepackage{tikz}
\usepackage{nicefrac}

\usepackage{changepage}
\usepackage{color, colortbl}

\usepackage{cleveref}

\usepackage{blindtext}
\usepackage{amssymb}

\DeclareMathAlphabet\mathbfcal{OMS}{cmsy}{b}{n}

\usepackage{nicefrac}
\usepackage{makecell}
\usepackage{multirow}
\usepackage{tabularx}
\usepackage{booktabs}
\usepackage{array}
\usepackage{longtable, tabularx}
\newcolumntype{L}[1]{>{\raggedright\let\newline\\\arraybackslash\hspace{0pt}}m{#1}}
\newcolumntype{C}[1]{>{\centering\let\newline\\\arraybackslash\hspace{0pt}}m{#1}}
\newcolumntype{R}[1]{>{\raggedleft\let\newline\\\arraybackslash\hspace{0pt}}m{#1}}
\newcolumntype{Y}{>{\centering\arraybackslash}X}
\newcolumntype{L}[1]{>{\raggedright\arraybackslash}p{#1}}

\newcolumntype{C}[1]{>{\centering\arraybackslash}p{#1}}

\newcolumntype{R}[1]{>{\raggedleft\arraybackslash}p{#1}}

\usepackage{filecontents}

\usepackage[normalem]{ulem}

\newacronym{isru}{ISRU}{\textit{in-situ} resource utilisation}
\newacronym{rlp}{rlp}{random loose packing}
\newacronym[plural=rcps,firstplural=random close packings (rcps)]{rcp}{rcp}{random close packing}
\newacronym{dfg}{DFG}{German Physics Foundation}
\newacronym{hp3}{HP3}{heat flow and physical properties package}
\newacronym{gtb}{GTB}{GraviTower Bremen}
\newacronym{mp}{MP}{Institute of Material Physics for Space}
\newacronym{elgra}{ELGRA}{European Low Gravity Research Association}
\newacronym{zarm}{ZARM}{Center of Applied Space Technology and Microgravity}
\newacronym{xct}{XCT}{X-ray computed tomography}
\newacronym{pmma}{PMMA}{polymethylmethacrylate}
\newacronym{1g}{$1g$}{gravity on-ground}
\newacronym{mug}{$\mu g$}{microgravity}
\newacronym{ptfe}{PTFE}{polytetrafluoroethylene}
\newacronym{ps}{PS}{polystyrene}
\newacronym{dem}{DEM}{discrete element methods}
\newacronym{iss}{ISS}{international space station}
\newacronym{eva}{EVA}{extravehicular activity}
\newacronym{vdw}{vdW}{van der Waals}

\bibliography{bibliography.bib}

\title{Granular jamming and rheology in microgravity}

\begin{document}

\newcommand{\red}[1]{\textcolor{black}{#1}}

\newcommand\blfootnote[1]{%
  \begingroup
  \renewcommand\thefootnote{}\footnote{#1}%
  \addtocounter{footnote}{-1}%
  \endgroup
}

\definecolor{brickred}{rgb}{0.8, 0.25, 0.33}
\definecolor{darkorange}{rgb}{1.0, 0.55, 0.0}
\definecolor{persiangreen}{rgb}{0.0, 0.65, 0.58}
\definecolor{persianindigo}{rgb}{0.2, 0.07, 0.48}
\definecolor{cadet}{rgb}{0.33, 0.41, 0.47}
\definecolor{turquoisegreen}{rgb}{0.63, 0.84, 0.71}
\definecolor{sandybrown}{rgb}{0.96, 0.64, 0.38}
\definecolor{blueblue}{rgb}{0.0, 0.2, 0.6}
\definecolor{ballblue}{rgb}{0.13, 0.67, 0.8}
\definecolor{greengreen}{rgb}{0.0, 0.5, 0.0}
\definecolor{razzmatazz}{rgb}{0.89, 0.15, 0.42}
\definecolor{ultramarine}{rgb}{0.07, 0.04, 0.56}
\definecolor{midnightgreen}{rgb}{0.0, 0.29, 0.33}
\definecolor{lavenderpurple}{rgb}{0.59, 0.48, 0.71}
\definecolor{bittersweet}{rgb}{1.0, 0.44, 0.37}
\definecolor{amaranth}{rgb}{0.9, 0.17, 0.31}
\definecolor{patriarch}{rgb}{0.5, 0.0, 0.5}
\definecolor{darkcandyapplered}{rgb}{0.64, 0.0, 0.0}
\definecolor{cadmiumgreen}{rgb}{0.0, 0.42, 0.24}
\definecolor{darkgreen}{rgb}{0.0, 0.2, 0.13}
\definecolor{deepcarrotorange}{rgb}{0.91, 0.41, 0.17}
\definecolor{deepcarmine}{rgb}{0.66, 0.13, 0.24}
\definecolor{maroon}{rgb}{0.69, 0.19, 0.38}
\definecolor{midnightblue}{rgb}{0.1, 0.1, 0.44}
\definecolor{lava}{rgb}{0.81, 0.06, 0.13}
\definecolor{ceruleanblue}{rgb}{0.16, 0.32, 0.75}
\definecolor{sacramentostategreen}{rgb}{0.0, 0.34, 0.25}
\definecolor{darkgreen}{rgb}{0.12, 0.3, 0.17}%
\definecolor{carmine}{rgb}{0.92, 0.3, 0.26}%
\begingroup
\sffamily
\author[1,2]{\sffamily Olfa D'Angelo\thanks{}}
\author[2,3,4]{\sffamily Qing Yu}
\author[1]{\sffamily Thorsten P\"{o}schel}

\affil[1]{\sffamily\scriptsize Institut Sup\'{e}rieur de l'A\'{e}ronautique et de l'Espace (ISAE-SUPAERO), Universit\'{e} de Toulouse, Toulouse, France}
\affil[2]{\sffamily\scriptsize Institute for Multiscale Simulation, Universit\"{a}t Erlangen-N\"{u}rnberg, Cauerstra\ss{}e 3, 91058 Erlangen, Germany}
\affil[3]{\sffamily\scriptsize Technical University of Munich (TUM), TUM School of Natural Sciences, Department of Chemistry, 85748 Garching, Germany}
\affil[4]{\sffamily\scriptsize Research Group Electromobility and Learning Systems, Technische Hochschule Ingolstadt, 85049, Germany}

\date{}

\twocolumn[
\begin{@twocolumnfalse}
  \maketitle

\vspace{-20pt}
\begin{abstract}\sffamily\noindent
Understanding how granular materials behave in low gravity is crucial for planetary science and space exploration. 
It can also help us understand granular phenomena usually hidden by gravity.
On Earth, gravity dominates granular behavior, but disentangling its role from intrinsic particle interactions is challenging. 
We present a series of compression and shear experiments conducted in microgravity using the \acrfull{zarm} drop tower and \acrfull{gtb}. 
Our in-house developed experimental setup enables precise measurement of packing density and \textit{in-situ} shear stress via a Taylor-Couette rheometer. 
We find that the jamming transition occurs at lower packing density in microgravity than on Earth, confirming that gravity promotes densification. 
Rheological measurements further reveal that in microgravity, the lack of a secondary force field and predominance of cohesive interparticle forces
increase the stress needed for granular media to flow.
These findings highlight gravity's dual role in enhancing both compaction and flow, and demonstrate the need for tailored granular models, valid in low- and microgravity environments.
\vspace{20pt}
\end{abstract}
\end{@twocolumnfalse}
]
\endgroup

\def\lengthl{35e-3} 
\def\radiusinternal{25e-3} 

\section{Introduction}
\blfootnote{\raggedright\noindent * Corresponding author:\\ Olfa D'Angelo, olfa.dangelo@fau.de}

From celestial bodies to fine powders, a wide array of materials are \emph{granular}: 
a collection of particles, too large for thermal fluctuations, too loosely bound together to form a solid. 
Because of the weakness of the bonds that link the particles together,
granular flows are defined by the interplay between these bonds and external loads --
among which is gravity. 
While granular media constitute countless human-made and natural materials,
predicting and controlling their behavior remains a challenge.
This increases hazards from natural phenomena (avalanches, pyroclastic flows)
and can be ecologically and economically damaging (e.g., sub-optimal efficiency of industrial powder handling processes).

We study the packing and flowing behavior of granular materials in microgravity. 
By understanding how granular media behave when gravitational acceleration is reduced, 
we can isolate the effect of particles' weight from 
intrinsic granular properties.
Freed from gravity, competing microscopic forces that govern granular matter can be highlighted, 
providing a ground-truth to model their flow behavior.
Primary activities of space exploration missions also rely on 
identifying the transition at which a granular material shifts behavior from solid-like to fluid-like (its un-jamming transition),
and once flowing, on predicting its rheology.
Typical applications in space include
building large structures and habitats,
processing regolith for \textit{in-situ} oxygen and metals production, 
or motorized exploration of asteroids and planetary surface.

Jamming describes the transition between the flowing state,
in which the granular fabric undergoes plastic deformation, 
and the jammed state, where a stable contact network can withstand a certain amount of stress without large-scale reorganization, akin to a solid~\cite{Behringer2019}. 
The jamming transition happens at a finite packing density, $\varphi_\text{J}$.
It is indicated by a sudden rise 
in the number of contacts per particle, and macroscopically,
a sudden rise of the material's effective viscosity.

The packing density at jamming has been shown to vary with the gravitational acceleration,
both in 2D~\cite{Dorbolo2011} and 3D~\cite{DAngelo2022}.
For the latter, experiments consisting of a piston rising through a granular medium, conducted on parabolic flights, showed 
that the jamming transition happens at lower density in microgravity than on-ground
under Earth gravitational acceleration, $g_E$:
$\varphi_\text{J}(g_\mu) < \varphi_\text{J}(g_E) $.

Granular rheology experiments have also been conducted in microgravity.
\citeauthor{Murdoch2013} \cite{Murdoch2013,Murdoch2013b}, 
using a Taylor-Couette geometry on parabolic flight (without stress measurement), 
found that the secondary radial displacement field is driven by a gravity-induced gradient,
hence disappearing in microgravity.
A similar setup including stress measurement was used by 
\citeauthor{Bossis2004}~\cite{Bossis2004} to study dilute to dense systems of iron beads (\SI{1}{\mm} and \SI{2}{\mm}),
confirming the classical scaling argument by \citeauthor{Bagnold1954}~\cite{Bagnold1954}, $\sigma\propto\dot\gamma^2$, but finding prefactors largely exceeding expectations.
Generally, 
rheological studies in microgravity have reported increased peak shear strength \cite{Sture1998} and friction angles \cite{Alshibli2000, Marshall2018, Karapiperis2020} compared to terrestrial conditions. 
Classical soil mechanics experiments adapted to partial- and microgravity
also find increased dilation as gravity decreases \cite{Marshall2018, Karapiperis2020}.

Smaller-scale experiments employing devices like hourglasses \cite{Dorbolo2013,Brucks2008,Reiss2014,Hofmeister2009, Ozaki2023, Madden2025, Gaida2025, DAngelo2025c}, rotating drums \cite{Klein1990, Brucks2007, Brucks2008, Kleinhans2011}, fluidized bed~\cite{Williams2008}, or avalanche setups \cite{Brucks2008}, 
consistently report lower packing densities and increased prominence of cohesive forces:
powders that flow readily under Earth gravity exhibit cohesive behavior in low gravity, leading to aggregation and cluster formation which hinder flow \cite{Williams2008, Elekes2021, Love2014}.

An antagonist effect has also been reported.
Lower gravitational acceleration reduces 
the confining pressure applied to a granular bed.
On a sandy soil, it means that at shallow depth, 
the uppermost layer of soil is a dust-like, easily deformable material:
the strength and soil's bearing capacity are reduced~\cite{Singh2015,Mo2021};
the first centimeters of penetration test require low effort~\cite{Slyuta2014, Featherstone2021},
and in avalanche flows, the fluid-like state can be prolonged~\cite{Shinbrot2004},
giving the impression of an easily flowing, fluid-like material.

This twofold effect complicates
predictions of granular behavior outside of Earth gravitational field,
which had severe consequences in the past.
During the Apollo 15 Moon mission, astronauts lost hours of \gls{eva} struggling to insert a core sampling penetrator into the lunar soil \cite{Apollo15TechReport}. 
In 2009, the Mars rover Spirit
was abandoned after six years of loyal services
because it got embedded in soft granular soil, 
after all attempts at rescuing maneuvers failed~\cite{Kerr2009}.
More recently,
the \acrshort{hp3} Mole penetrator of
NASA's InSight Mars lander
failed to penetrate the Martian soil below \SI{37}{\centi\meter}~\cite{Spohn2022},
a much shallower depth than planned.
In future crewed missions, such mis-prediction could have catastrophic effects,
notably risking astronauts' safety. 

We present a series of experiments in microgravity, 
where the effective gravitational acceleration, $g$, tends to zero, $g_{\mu } := g \to 0$.
We investigate experimentally the influence of gravitational acceleration on:
\begin{enumerate}[label=(\roman*), topsep=0pt,parsep=0pt, partopsep=0pt, noitemsep]
\item the packing fraction at which the granular jamming transition occurs;
\item the rheology of the granular media under shear.
\end{enumerate}

We use a simple experimental setup
to first fluidize, then compress a granular packing until jamming. 
The packing density at jamming, $\varphi_\text{J}$, is measured
at a given gravitational acceleration, $g$.
Then, the granular material is subjected to shear, also under the given $g$,
while measuring shear stress, $\tau$, and variations in normal stress, $\Delta\sigma$.
The same experiments are reproduced in microgravity and on-ground ($g_E$).
We access microgravity using the two types of drop towers 
available at the \gls{zarm} (Bremen, Germany),
namely, free fall with catapult system \cite{vonKampen2006} and active (guided) \cite{Konemann2015,Gierse2017,Gierse2022} drop towers.

We first describe the methods, notably focusing on the in-house developed 
experimental setup and the platforms used to conduct microgravity experiments.
We then present our results on the 
packing density at jamming.
Using fluidization to reinitialize our packing, we also observe the effect of gravity on the fluidization behavior.
We then present our results on the rheology of granular materials in microgravity. 
Finally, we discuss how both changes in $\varphi_\text{J}$ and granular rheology observed in microgravity 
can be understood in a common framework.

\section{Methods}\label{sec:methods}

\subsection{Experimental procedure}
\begin{figure}[h!]
\centering
\includegraphics{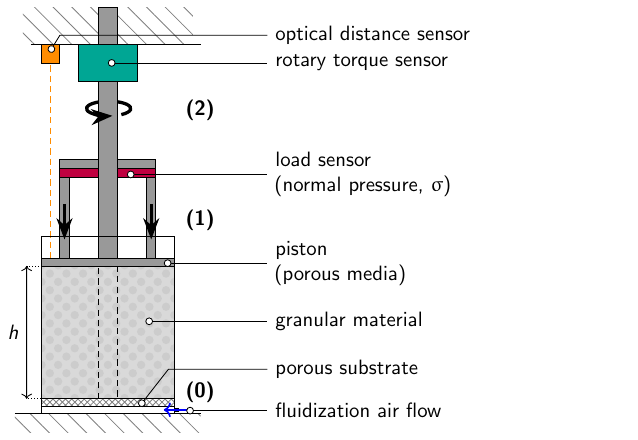}
\caption{\label{fig:schematic}
\textbf{Schematic of experimental setup.}
The basic experiments conducted consist of (1)~compression and (2)~shear of a granular material.
Data recording devices are:
an optical distance sensor (orange) to measure the piston's height, $ h $;
a rotary torque sensor (green) to measure shear stress, $\tau$;
a normal load sensor (red) to estimate normal pressure changes, $\Delta\sigma$, due to the material's expansion or densification.
}
\end{figure}

The experimental procedure consists of the following (\textit{cf.}\ Figure~\ref{fig:schematic}):
\begin{enumerate}[label=(\arabic*), start=0] 
\item Before the start of the experiment, the material is fluidized to reinitialize a reproducible, low density packing.

\item The material is compressed by a piston 
until the pressure reaches the value $p_\text{J}$.
The exact position of the piston is measured by an optical distance sensor, and used to calculate the volume occupied by the material,
and in turn the packing fraction at jamming, $\varphi_\text{J}(g)$.

\item Once compressed, 
the mechanical (rheological) properties of the jammed material are probed by imposing shear through rotation of the inner cylinder.
The piston is not at fixed position but can move slightly up and down
to accommodate for packing reorganization, while imposing a normal stress.
\end{enumerate}

\begin{figure}[h!]
	\centering
\includegraphics{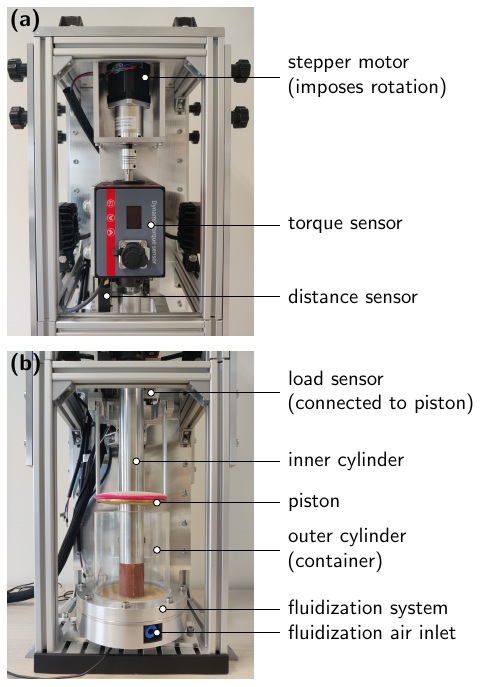}
	\caption{
		\textbf{Picture of experimental setup.} (a)~Upper and (b)~bottom part of the setup.
The full setup is placed inside a capsule to undergo microgravity experiments, in the drop tower or in the \gls{gtb}. 
	}
	\label{fig:photosetup}
\end{figure}

The experimental setup is shown in Figure~\ref{fig:photosetup}. 
It comprises three main components: 
a fluidization system,
a ring-shaped piston for compression,
and a Taylor-Couette geometry rheometer (two coaxial cylinders with the inner cylinder rotating).
The setup can slide opened to be divided into two parts, allowing for easy change of the granular material.

The piston is controlled by a linear actuator 
(BE069-3 from \textit{Befenybay}),
and connected to a normal load sensor.
Values from the load cell (in \si{\kg}) are converted into normal pressures, $\sigma$.

Once in contact with the powder bed, the piston position, $h$, gives the volume occupied by the material,
from which we calculate its global packing fraction.
The piston position is recorded by a triangulation laser distance sensor (OMom30-L0100.HV.TXN, Baumer),
which has a precision of \SI{1}{\micro\meter}.
This sensor is selected based on 
the previously identified difference between $\varphi_\text{J}$ on Earth and in microgravity~\cite{DAngelo2022},
which, applied to our setup's scale, is expected to be approximately \SI{0.1}{\mm}.

During the compression test (measurement (1)),
the jamming point is indicated by the halt of the piston and sudden rise of pressure. 
The vertical pressure at which the piston is stopped is empirically fixed at $ p_\text{J} = \SI{888}{\pascal}$.
Note that for our purpose -- comparing the packing fraction at which jamming occurs under Earth and microgravity conditions -- the
exact pressure value is less critical than ensuring consistency in the experimental procedure across both environments. To account for the piston assembly's own weight, the measured load at time $t = 0$ is used as an offset (corresponding to the start of descending of the piston, at the same $g$ as the experiment will take place).

To ensure quasi-static compression,
the piston speed is fast at first (\SI{2.5}{\milli\meter\per\second}),
then becomes slower as the pressure increases.
It slows down to \SI{0.75}{\milli\meter\per\second} once the
normal pressure reaches \SI{674}{\pascal}.
These values are determined empirically.

During the subsequent shear test (measurement (2)), 
the piston has the freedom to slightly move up and down to give space to the granular material to reorganize.
This is implemented by a spring linking the piston to the linear actuator (see Fig.~\ref{fig:pistonsystem}).
A load cell is placed between the actuator (fixed during measurement (2)) and the spring, 
to capture compaction or dilation of the granular medium during shear. 
Variations in normal pressure, $\Delta\sigma$, are given in regard to the normal pressure 
at the onset of shear.

\subsection{Microgravity platforms}

Microgravity conditions ($g_{\mu}$) -- that is, effective weightlessness obtained by free fall under Earth gravity -- are provided by the \gls{zarm} drop towers in Bremen (Germany).
Two different systems are used:
\begin{itemize}
\item The \acrfull{gtb}, a relatively short active drop tower \cite{Konemann2015,Gierse2017}. In the \gls{gtb}, the experimental capsule is actively guided to match the speed variations of free fall. It provides \SI{2.3}{\second} of microgravity. 
\item The (passive) drop tower with catapult system, where the experiment describes a parabola in vacuum, providing \SI{9.3}{\second} of effective microgravity \cite{vonKampen2006}.
\end{itemize}

\begin{figure}[htb!]
\centering
\includegraphics{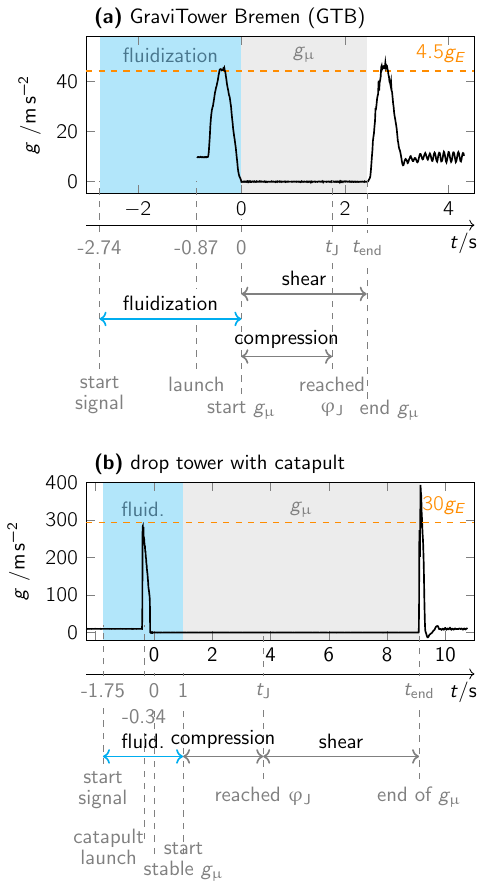}
\caption{ \label{fig:acceleration-profiles}
\textbf{Acceleration in the experimental capsule and corresponding experiment timeline.} 
(a) Acceleration \textit{versus} time during a \gls{gtb} experiment. 
Shear and compression experiments are conducted independently: either one or the other is conducted during each experiment.
(b) Acceleration during a drop tower experiment with catapult. Compression, then shear are conducted consecutively.
Note that the instant at which the jamming point is reached, $t_\text{J}$, 
is not time-controlled but controlled by the pressure imposed by the piston.
}
\end{figure}

Typical acceleration profiles measured 
on either of the platforms are provided in Figure~\ref{fig:acceleration-profiles}.
We also provide the nominal timing for all types of experiment conducted.

Data collection in microgravity is constrained by the limited availability of experimental platforms.
For reference, each experiment is also performed under Earth gravity ($g_{\text{E}}$); 
ground repetitions follow a microgravity run as closely as possible, to ensure the same conditions.

\subsection{Materials}\label{sec:materials}

Two granular materials are used in our experiments,
the first one constituted of
monodispersed polystyrene spherical particles of diameter $d = \SI{1}{\mm}$
(polystyrene bulk density $\rho_\textsc{PS} = \SI{1.05}{\g\per\cm\cubed}$),
with relatively smooth surface. 
Similar particles 
(polystyrene, monodispersed, spherical) form the second material,
but with much smaller diameter, $d =\SI{80}{\micro\m}$.
They also have a relatively smooth surface.
Both materials are manufactured by \textit{Dynoseeds} under the commercial name \textit{Microbeads}; 
their surface state and flow-behavior are characterized in Refs.~\cite{DAngelo2021,DAngelo2025}.

By using two different particle sizes, we vary the interparticle cohesion.
Cohesion arises from attractive forces between particles, originating from mechanisms such as \gls{vdw} interactions, electrostatic forces, or capillary bridges, among others. These cohesive forces add a new force scale -- hence a new time scale -- to granular flows \cite{Pouliquen2025}.
Both materials are considered dry: polystyrene particles are hydrophobic, and
besides standard precautions (storage under dry conditions, anti-humidity agents inside storage containers, limiting exposure outside containment), 
dry air is used for fluidization.
For dry, uncharged particles, \gls{vdw} forces dominate \cite{Affleck2023}. Here, we exploit the size-dependence of \gls{vdw} interactions to compare granular materials that might exhibit different cohesion, albeit being otherwise identical.

To estimate the \gls{vdw} force, $F_\text{vdW}$, we consider two spherical particles of equal radius $r$,
whose surfaces are separated by a distance $\ell$:
\begin{equation}\label{eq:vdWF}
F_\text{vdW} = \frac{H}{12} \frac{r}{\ell ^2},
\end{equation}
where $H=\SI{e-19}{\joule}$ is the Hamaker coefficient,
a material-dependent constant \cite{Hamaker1937, Israelachvili2011} (value for for polystyrene). 
We assume $\ell = \mathcal{O}(\SI{10}{\nano\meter})$ \cite{Krupp1966,Affleck2023},
noting that while numerical values are sensitive to $\ell$, the overall scaling behavior remains the same.

From Eq.\,\ref{eq:vdWF}, $F_\text{vdW}$ scales linearly with particle size, $F_\text{vdW}\propto r$.
By contrast, inertial forces such as particle weight scale with volume, $F_\text{g} \propto r^3$. 
Consequently, \gls{vdw} forces dominate for very fine powders, even though their absolute magnitude is smaller than for larger particles of the same material.

\begin{figure}
\includegraphics{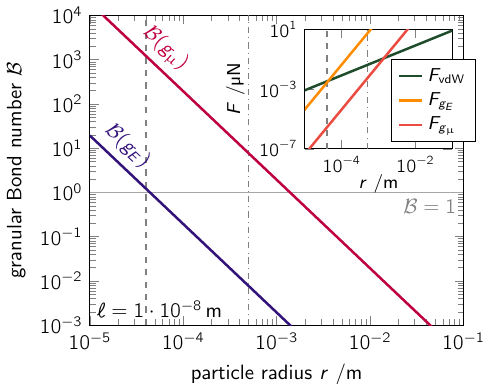}
\caption{\label{fig:cohesionvasize}
\textbf{Granular Bond number, $\mathbfcal{B}$, for increasing particle size. }
$\mathcal{B}$ is calculated, as defined in Eq.~\ref{eq:mynameisbond},
for $g_E=\SI{9.81}{\m\per\square\second}$ and for $g_\mu =10^{-3} g_E$.
Vertical grey lines indicate the tested particle diameters (dashed: $d=\SI{80}{\micro\meter}$; dash-dotted: $d=\SI{1}{\mm}$);
a horizontal line at $\mathcal{B}=1$ marks the transition to cohesive behavior (above).
The inset shows the \gls{vdw} force $F_\text{vdW}(r)$, and $F_\text{g}(r)$ for both $g_E$ and $g_\mu $. 
The main panel and inset have the same $x$-axis boundaries.
}
\end{figure}

This competition is quantified by the granular Bond number, $\mathcal B$, which measures the ratio of cohesion to inertia \cite{Affleck2023, Gaida2025}. Approximating $\mathcal B$ as the ratio of \gls{vdw} forces to particle weight, we obtain:
\begin{equation}\label{eq:mynameisbond}
\mathcal B = \frac{F_\text{vdW}}{F_\text{g}} = \frac{H}{16 g \rho \pi  }\, \frac{1}{  \ell^2  \, r^2}
\end{equation}
for particles of density $\rho$ under gravitational acceleration $g$.
Figure~\ref{fig:cohesionvasize} shows $\mathcal B (r)$, as defined above, for $g_E$ and $g_\mu =10^{-3} g_E$.
The inset compares directly $F_\text{vdW}(r)$ and $F_{g}(r)$ for both $g_E$ and $g_\mu$. 

The transition to cohesive behavior, which should happen at $\mathcal{B}=1$, 
is hence expected to occur around our smaller particle size under Earth gravity, 
and around our larger particle size in microgravity.

\subsection{Air-fluidization}

\begin{figure}[h]
	\centering
\includegraphics{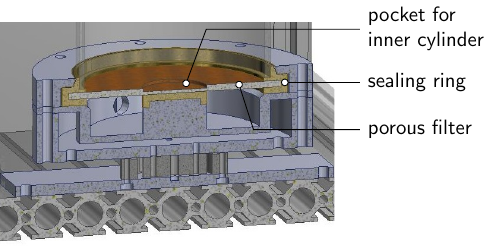}
	\caption{\label{fig:fluidization-profile}
\textbf{Section view of the fluidization system.} 
This assembly is placed at the bottom the experimental setup, 
as indicated in Fig.\,\ref{fig:photosetup}.
	}
\end{figure}

An air-fluidization system is installed to erase the stress history of the granular media before each experiment.
Figure~\ref{fig:fluidization-profile} shows a section view of the fluidization system.

A sealing ring is held in place by two metal pieces screwed together,
holding the container cylinder and a porous filter with
\SI{35}{\micro\meter} pore size;
it creates a homogeneous gas flow in the container for uniform fluidization.
To ensure proper centering of the inner cylinder, a cylindrical pocket is carved into the filter,
fitted to the inner cylinder diameter.

\subsection{Piston and packing fraction at jamming}\label{sec:piston}

\begin{figure}[h!]
	\centering
\includegraphics{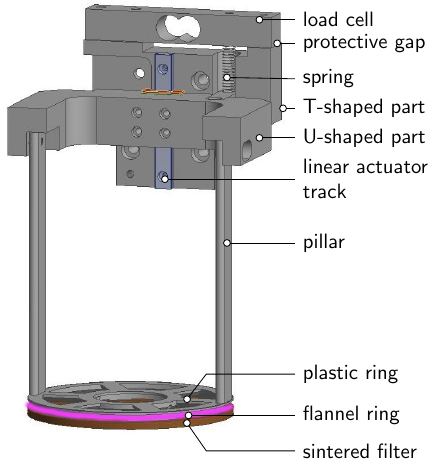}
	\caption{\label{fig:pistonsystem}
\textbf{Piston assembly.} The T-shaped structure protects the load cell from hypergravity-induced overload. The U-shaped connector transmits force between the load cell and the piston through a spring. The piston features a layered sealing system: a sintered copper filter allows airflow, a flannel ring prevents particle leakage, and a plastic ring maintains alignment.
	}
\end{figure}

The piston assembly, as shown in Figure~\ref{fig:pistonsystem}, fulfills three functions: 
compressing the granular material into a jammed state;
ensuring reliable particle containment during fluidization and launch, while letting the air-flow out;
protecting the load cell from overload during hypergravity phases.
A T-shaped component, mounted on the linear actuator, shields the load cell from excessive force by maintaining a $\sim \SI{0.1}{\mm}$ gap below the load cell, where it is attached to the spring
(see \enquote{protective gap} in Fig.~\ref{fig:pistonsystem}).
A U-shaped component connects the piston to the load cell via a spring.
To prevent particle leakage, the piston itself features a sandwich structure: a sintered copper filter at the base permits airflow during fluidization, while a flannel ring, enclosed by a plastic ring, enhances sealing efficiency.

The packing fraction is calculated via 
\begin{eqnarray}
	\label{PFequ}
	\varphi_{J}=\frac{4 m}{{\rho_b}{\pi}(d_{c}^{2}-d_{i}^{2})h}
\end{eqnarray}
where $ m $ is the mass of the granular sample;
$\rho_b$ is the bulk density of the material composing the granular particles;
$d_c = \SI{84}{\mm}$ is the diameter of the container (outer) cylinder, and $d_i = \SI{30}{\mm}$ is the diameter of the inner cylinder.
$h$ is the piston position; it is calculated taking as reference the bottom of the container, such that it represents the height occupied by the granular material.

\subsection{Rheology}
The second phase of the experiment consists of probing the mechanical (rheological) properties of the material. 
This is achieved by applying shear through the rotation of the inner cylinder (see Figs.~\ref{fig:schematic} and \ref{fig:photosetup}).
The piston is not moved up before the onset of shear, but
the setup is designed such that the piston, connected to the structure via two springs, can move slightly up and down
(\textit{cf.}\ Sec.~\ref{sec:methods}.\ref{sec:piston} and Fig.~\ref{fig:pistonsystem}).

The inner cylinder rotation is controlled independently by a stepper motor (17HS19-1684S-PG100 from \textit{Stepper-Online}) 
of maximum torque \SI{4}{\newton\meter}.
The rotational stepper motor features a 100:1 gearbox to ensure ample torque provision.

The inner cylinder is rotated at constant speed of 
\SI{4.3}{\rpm}
in every rheology experiment presented.
Translated to shear rate (assuming Newtonian behavior), the material is sheared at $\dot\gamma=\SI{e-1}{\per\s }$.

The rotational torque sensor (ATO-TQS-DYN-200 from ATO) 
used to measure torque, $M$,
has a capacity range of $\SIrange{-0.1}{0.1}{\newton\meter}$,
with a precision of \SI{100}{\micro\newton\meter} and accuracy of \SI{30}{\micro\newton\meter}.
Sandpaper is used at the bottom of the inner cylinder to avoid slippage (coarse grit 60, typical rugosity $\sim \SI{250}{\micro\meter}$);
it covers a height $L=\SI{35}{mm}$.

The shear stress, $\tau$, is calculated from the recorded torque, $M$, as
\begin{equation}
\tau = \frac{M }{2 \pi \, h \, r_i}.
\end{equation}

The variations in normal pressure from that at the onset of shear, $\Delta\sigma$, are recorded throughout each shear test.

\subsection{Experimental uncertainty}

The experimental uncertainty of the packing fraction
determines whether the conclusions from our experiments are statistically relevant.
Each variable in Eq.~\ref{PFequ}
is measured independently; absolute uncertainties are given in Table~\ref{tab:abunc}.
The global experimental uncertainty of $\varphi_{J}$ is calculated according to the simplified propagation of uncertainty \cite{Ku1966}.
The maximum absolute uncertainty is found to be 0.0061, corresponding to a relative uncertainty of 1\%.

\begin{table}[h]
\sisetup{detect-all}
\normalfont\sffamily\small
\sansmath\sffamily\small
\centering
\caption{\label{tab:abunc}
\textbf{Absolute uncertainty on each experimental variable.} The values recorded here are used in the calculation of the experimental uncertainty.
}
\renewcommand{\arraystretch}{1.05}
	\begin{tabular}{c l c c c}
	\toprule
	\multicolumn{2}{c}{\textbf{variable}} & \textbf{unit} & \makecell{\textbf{absolute}\\\textbf{uncertainty}} \\
	\midrule
        $\rho_b$ &\thead[l]{material\\bulk density}&\si{\kg\per\m\cubed} &10\\ 
        $d_c$& \thead[l]{container\\outer diameter} & \si{\m} & \num{e-4}\\ 
		$d_i$& \thead[l]{inner cylinder\\diameter} &\si{\m}&\num{e-4}\\
        $h$& \thead[l]{piston position} &\si{\m}&\num{e-6}\\
        $ m $& \thead[l]{material\\sample mass} & \si{\kg} & $7\cdot\num{e-6}$\\
	\bottomrule
	\end{tabular}
\end{table}

\section{Results}

\subsection{Jamming transition density}

The compression experiment was conducted three times in microgravity using the \gls{gtb},
with the \SI{1}{\mm} diameter particles. 

\begin{figure}[htb!]
\centering
\includegraphics{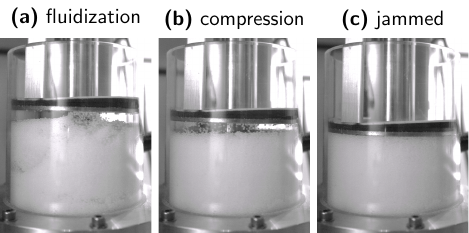}
\caption{\label{fig:fluidization-screenshots-gtb}
\textbf{Screenshots from the control video during \gls{gtb} experiment on
\SI{1}{\mm} diameter polystyrene particles.}
After (a)~fluidization, (b)~the piston is lowered to compress the material until measuring a steep increase in normal pressure up to
$p_\text{J}$, at which point (c)~the piston is stopped as the material is considered jammed.
Multimedia available online.
}
\end{figure}

\begin{figure*}[h!]
\centering
\includegraphics{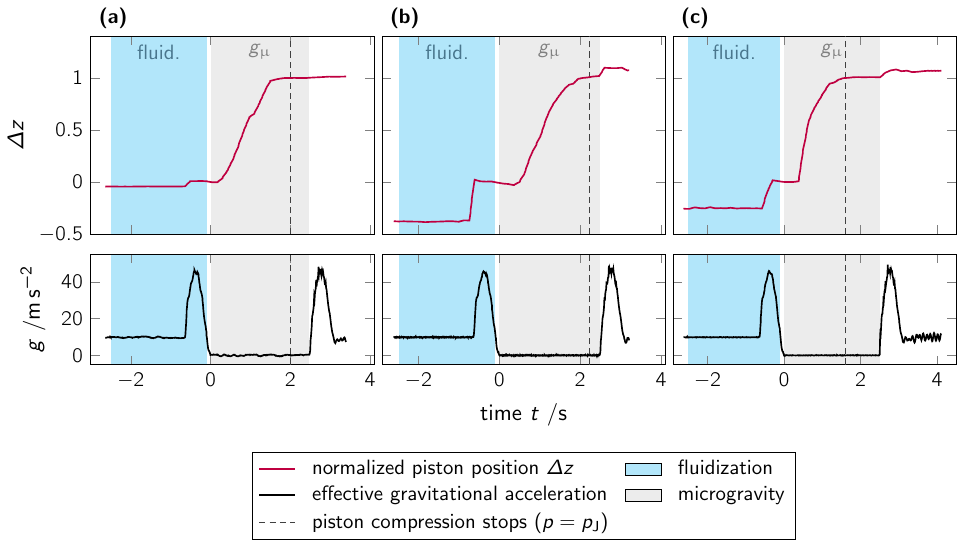}
\caption{\label{fig:phi-j-microgravity}
\textbf{Normalized piston position variation up to the jamming point and effective gravitational acceleration during three \gls{gtb} tests.}
The piston position, $\Delta z$, is normalized by its initial and final positions;
$g$ is the effective gravitational acceleration throughout a \gls{gtb} test.
All experiments shown here are conducted on the \gls{gtb} with $d=\SI{1}{\mm}$ spherical polystyrene particles.
The piston stops when reaching $p_\text{J}$, indicated by a dashed line.
Fluidization and microgravity and marked by shaded background (blue and grey, respectively).
Multimedia available online.
}
\end{figure*}

Snapshots from a representative experiment are shown in Figure~\ref{fig:fluidization-screenshots-gtb} (see video in online multimedia).
Three independent repetitions are presented in Figure~\ref{fig:phi-j-microgravity}(a-c).
Fluidization ceases as gravity approaches $g_{\mu}$, at which point the piston starts its downward motion for compression.

A short waiting time is introduced after fluidization to allow the packing to relax. At the onset of microgravity, a brief negative acceleration typically occurs, causing expansion of the granular packing and upward motion of some particles 
(visible in multimedia video available online; see also Fig.~\ref{fig:fluidization-screenshots-gtb}(b)).

\begin{figure}[h!]
\centering
\includegraphics{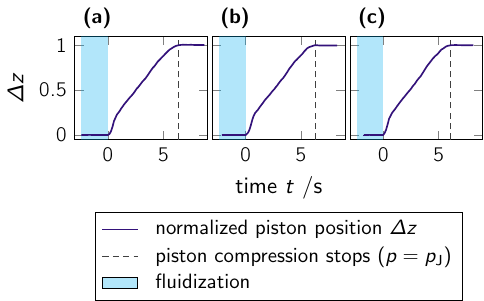}
\caption{\label{fig:phi-j-ground}
\textbf{Normalized piston position variation up to the jamming point during tests in Earth gravity.}
The piston position, $\Delta z$, is normalized by its initial and final positions.
All experiments shown here are conducted with $d=\SI{1}{\mm}$ spherical polystyrene particles (note that three out of ten repetitions are shown here).
The piston stops when reaching $p_\text{J}$, indicated by a dashed line.
Fluidization is marked by blue shaded background.
}
\end{figure}

\begin{figure*}[h!]
\includegraphics{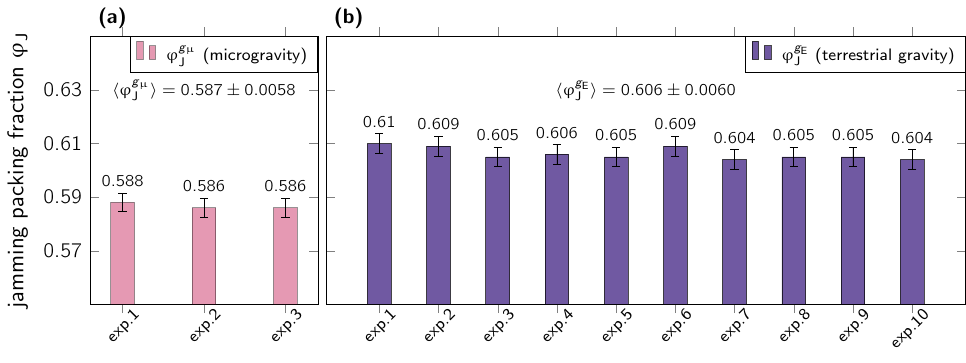}
    \centering
	\caption{\label{fig:GTBPF}
\textbf{Global packing fraction at jamming, $\varphi_\text{J}$, for all microgravity and Earth gravity compression tests conducted.}
All experiments shown here are conducted with $d=\SI{1}{\mm}$ spherical polystyrene particles;
all microgravity experiments were conducted in the \gls{gtb}.
Error bars are the experimental uncertainty on the measurement. 
$\langle \varphi_\text{J} \rangle$ are averages over each repetitions, given with the variance over all repetitions per $g$. 
	}
\end{figure*}

For the sake of comparison, three exemple datasets in Earth gravity, $g_\text{E}$, are provided in Fig.~\ref{fig:phi-j-ground}. 
A total of ten independent repetitions of the experiment were conducted, as reported in Fig.~\ref{fig:GTBPF}(b).

In Figures~\ref{fig:phi-j-microgravity} and \ref{fig:phi-j-ground},
the piston's vertical position is normalized by its initial and final position, 
so that $\Delta z = 0$ is its starting position and $\Delta z =1$
the position at which the packing jams;
i.e., let $h_0$ be the starting height of the piston
and $h_\text{J}$ the height at which the packing jams, 
$\Delta z (h) = (h-h_0 )/(h_\text{J} - h_0)$.
Time is expressed either from the start of microgravity (Fig.~\ref{fig:phi-j-microgravity}), or from the onset of piston's motion for ground experiments (Fig.~\ref{fig:phi-j-ground}).

A summary of the packing fractions measured in all experiments is presented in Fig.~\ref{fig:GTBPF}(a,b), corresponding to microgravity and ground-based conditions, respectively. 
Error bars indicate the experimental uncertainty associated with the measurement.

From Figure~\ref{fig:GTBPF}, we see that the jamming packing fraction, $\varphi_\text{J}$, is lower in microgravity than in terrestrial gravity on every repetition, with average values of 
$\langle \varphi_\text{J}^{g_\mu} \rangle = 0.587 $ and $\langle \varphi_\text{J}^{g_E} \rangle = 0.606 $.

To quantify variability within each dataset despite differing sample sizes, 
we calculate the coefficient of variation, $c_v$, defined as the ratio of the standard deviation to the mean
(also known as relative standard deviation).
We find $c_v^{g_\mu} = 0.1\%$ and $c_v^{g_E} = 0.4 \%$, indicating a larger dispersion in the measurements conducted under Earth gravity compared to microgravity, 
notwithstanding the unequal sample size
and generally low variability.

\subsection{Fluidization behavior}

We use fluidization to reinitialize the granular packing before each repetition, erasing its stress history
to ensure consistent initial conditions. 
Previous work by \citeauthor{Born2017} \cite{Born2017} demonstrated
that fluidization in microgravity 
leads to a homogeneous, fully mobilized state, in contrast to behavior observed under Earth gravity.

In the experiment described above (\SI{1}{\mm} polystyrene beads fluidized during a \gls{gtb} experiment, \textit{cf.}\ Fig.~\ref{fig:phi-j-microgravity}; multimedia available online),
the launch phase induces an acceleration of approximately $4.5 g_E$.
Fluidization is maintained throughout hypergravity and ceases at the onset of microgravity. 
Under hypergravity, the increased weight of the granular bed prevents full fluidization, 
resulting in an expanded but not fully agitated state.
A slight negative acceleration immediately following the hypergravity phase further aids the granular bed expansion before the start of the experiment.

\begin{figure*}[htb!]
\centering
\includegraphics{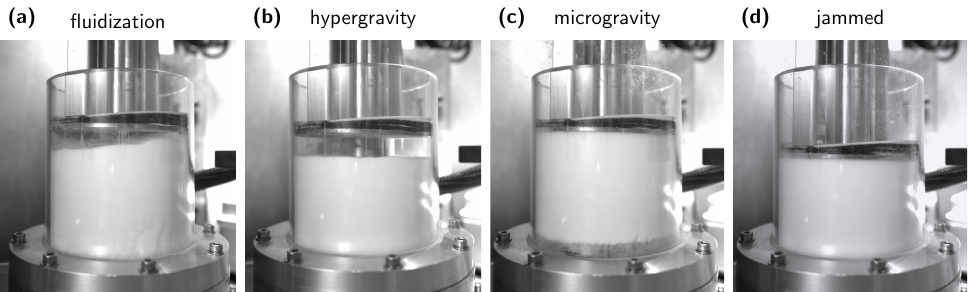}
\caption{\label{fig:fluidization-screenshots-droptower}
\textbf{Screenshots from the catapult drop tower experiment.} 
This experiment used \SI{80}{\micro\meter} polystyrene particles. 
The granular bed is first fluidized (a), then it experiences an initial compaction due to launch-induced hypergravity ($\sim 30 g_E$) (b); the fluidization is continued throughout hypergravity and for one second in microgravity. 
Once in microgravity, the entire bed is shifted upward as a block by the airflow (c).
The piston then starts descending to compress the granular packing (d).
}
\end{figure*}

Here, we focus on the second experiment, which uses much smaller \SI{80}{\micro\meter} polystyrene beads,
in the drop tower with catapult. This exposes the capsule to an initial acceleration of almost \SI{300}{\meter\per\square\second} ($\sim 30 g_E$, \textit{cf.}\ Fig.~\ref{fig:acceleration-profiles}(b)), compacting the granular bed significantly under its own weight. 
Before, during and for one second after this initial acceleration peak, 
the granular media is subjected to a fluidizing airflow at constant pressure $p = \SI{0.1}{\bar}$ -- identical to that used in ground-based fluidization.
Screenshots of the key phases, captured by the control camera, are shown in Figure~\ref{fig:fluidization-screenshots-droptower}.

Note that in both experiments (\gls{gtb} and drop tower), the same fluidization pressure is applied across all gravity regimes (Earth, hypergravity, microgravity). The purpose of fluidization in these experiments is not to investigate reduced-gravity fluidization \textit{per se}, 
but to prevent packing compression by reinitializing the bed as acceleration decreased following the launch hypergravity phase.

As seen in Fig.~\ref{fig:fluidization-screenshots-droptower}(c), just after hypergravity and while the airflow persists in microgravity, the granular material does not exhibit individual particle agitation. 
Instead, the entire granular bed is displaced as a cohesive block, 
behaving more like a porous solid than a fluidized discrete medium. 
When compression starts, this block is pushed downward uniformly, without visible signs of particle motion.

Packing fraction data from this experiment are excluded from further analysis, as the observed compaction could not be attributed to the controlled piston compression alone, but rather to the substantial pre-compaction induced by launch acceleration.

Nonetheless, this experiment illustrates the challenge of fluidizing highly cohesive powders in microgravity. 
Under Earth gravity, a balance between particle weight and upward airflow drag allows for a fluid-like, agitated state~\cite{Born2017,DAngelo2025a}.
In microgravity, however, the particles' weight being eliminated and cohesive forces dominating
can reduce the effectiveness of air-fluidization in agitating the granular material,
 pointing to the need for further investigation into mechanisms of granular bed reinitialization in low- and microgravity.

These findings point to the need for further investigation into mechanisms of granular bed reinitialization in low- and microgravity environments. 
For long-duration experiments on orbital platforms (e.g., the \gls{iss} and its successors), 
if air-fluidization is used for packing reinitialization, 
it should be supplemented with mechanical agitation to ensure effective disordering and reproducibility of the initial state.

\subsection{Rheology}

\begin{figure*}[htb!]
\includegraphics{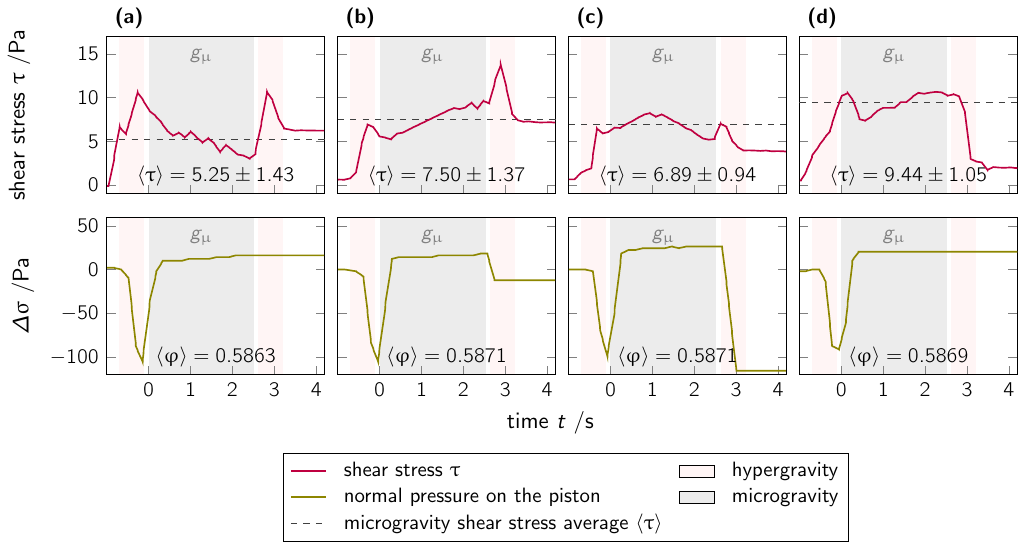}
	\caption{\label{fig:GTBPT}
\textbf{Shear stress $\tau$ and variations in normal pressure $\Delta \sigma$ for microgravity experiments conducted in the \gls{gtb}.}
All experiments shown here are conducted with \SI{80}{\micro\meter} diameter polystyrene particles.
Microgravity is marked by grey and hypergravity by red shaded background.
Dashed lines mark the average shear stress $\langle\tau\rangle$ in microgravity (numerical value and standard deviation for each measurement given in panel).
The average packing density during the test, $\langle\phi\rangle$, are given in the bottom panels (also compiled on the microgravity period). 
	}
\end{figure*}

\begin{figure*}[htb!]
\includegraphics{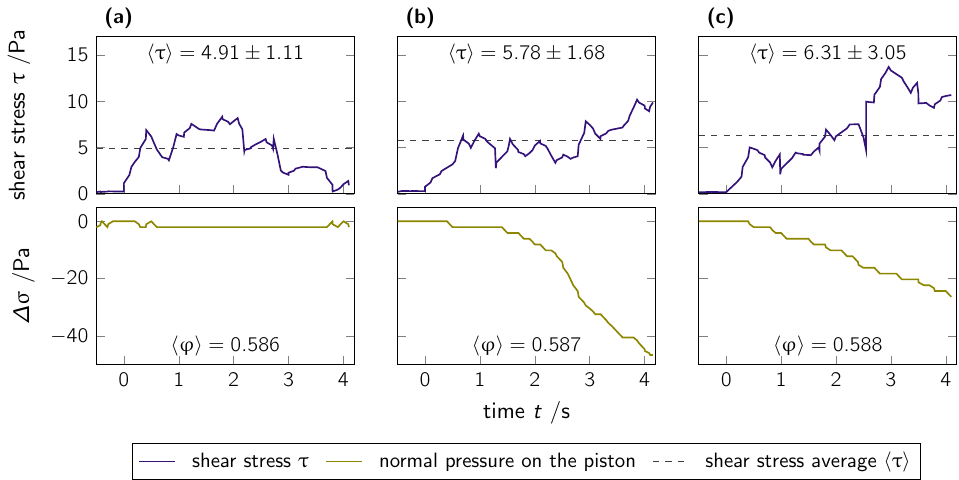}
	\caption{\label{fig:GTBPgroundT}
\textbf{Shear tests conducted under Earth gravity with \SI{80}{\micro\meter} diameter polystyrene powder.}
Graphs (a), (b) and (c) represent three repetitions of the same experiment.
The upper panels show the shear stress \textit{versus} time, as well as the average shear stress over the test, $\langle \tau \rangle$ (plotted as dashed line).
The lower panels show the changes in normal pressure, $\Delta\sigma$, from the onset of shear (at $t=0$).
$\langle \varphi \rangle$ is the average packing fraction over the test.
	}
\end{figure*}

\begin{figure*}[htb!]
\includegraphics{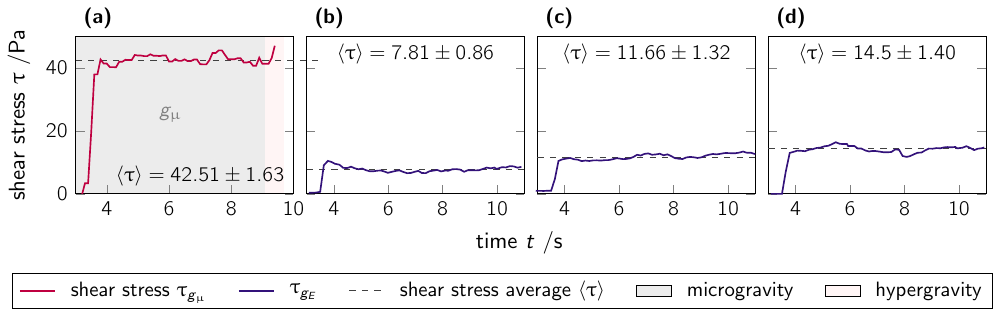}
	\caption{\label{fig:dtbtorque}
\textbf{Shear stress \textit{versus} time for drop tower shear test and corresponding individual ground test, for polystyrene particles of diameter $\bf{d=}$\,\SI{1}{\mm}.}
Panel (a) shows drop tower test data; the light grey shading indicates microgravity, light red shading hypergravity. 
Panels (b), (c) and (d) show the corresponding data for terrestrial gravity. 
The dashed lines show the time averaged data, $\langle \tau \rangle$.
	}
\end{figure*}

Two types of shear tests were conducted.
First, four repetitions of shear experiments 
using fine polystyrene powder ($d = \SI{80}{\micro\meter}$),
performed in the \gls{gtb}.
The results are shown in Figures~\ref{fig:GTBPT} for microgravity and \ref{fig:GTBPgroundT} for the corresponding ground-based experiments. 
Second, an additional shear test was conducted in the drop tower with a catapult system with large polystyrene beads ($d=\SI{1}{\mm}$) (Figure~\ref{fig:dtbtorque}(a)),
and was subsequently reproduced on-ground (Figures~\ref{fig:dtbtorque}(b-d)).

In both experimental configurations, the duration of microgravity -- limited to seconds -- does not allow the system to reach a steady state. 
In sheared granular materials, the relaxation times can be long, with steady state sometimes requiring hours to establish, notably in Taylor-Couette shear geometry~\cite{DAngelo2025a}. 
The microgravity platforms used here
inherently constrain the duration of experiments, and are thus not suited for long-term behavior.

Figure~\ref{fig:GTBPT} shows the results of the four shear tests conducted in the \gls{gtb}. 
Each panel shows the shear stress, $\tau$ (upper plot) and the corresponding change in normal pressure, $\Delta\sigma$ (lower plot),
relative to the baseline measured immediately before the start of the shear experiment
(typically at $t = \SI{-1}{\s}$). 
The normal pressure is not controlled but only measured; 
it provides information on wether the granular bed expands or compacts during shear
(remember that the piston is mounted on a spring, \textit{cf.}\ Figure~\ref{fig:pistonsystem}).
The onset of inner cylinder rotation happens at $t = \SI{-1}{\s}$;
microgravity starts at $t=0$. The material is sheared throughout the hypergravity phase,
such that the shear stress recorded corresponds to the stress needed to maintain and not initiate shear in microgravity.
The shaded grey area denotes microgravity; shaded red, hypergravity.

Shear stress measurements across the four microgravity experiments are relatively consistent,
with an average shear stress over all repetitions of $\langle \tau \rangle_{g_\mu } = \SI{7.27}{\pascal}$ ($c_v^{g_\mu}=24\%$).
Temporal fluctuations within individual experiments account for coefficients of variations between 
11\% and 27\%,
with no systematic trend of increase or decrease.

The normal pressure variations, $\Delta\sigma$,
exhibits a sharp drop during the hypergravity phase prior to microgravity onset, reflecting the strong compaction of the granular bed under the combined influence of downward acceleration and shear. 
This is followed by a sudden expansion upon entering microgravity, leading to a consistent increase in normal pressure of approximately $+\SI{20}{\pascal}$ across all repetitions.

These results can be compared to the corresponding
experiments in terrestrial gravity (Figure~\ref{fig:GTBPgroundT}).
\red{
In microgravity, we initiated shearing \emph{before} the onset of microgravity, 
to minimise start-up transients and obtain values closer to steady state, despite the short duration at $g_\mu $
(i.e., shearing starts at $t<0$, while we compute $\langle \tau \rangle $ only \emph{during} microgravity; see Fig.~\ref{fig:GTBPT}).
To avoid biasing the ground data toward lower values, 
we compute the ground averages starting at $t=\SI{0.5}{\s}$,
and for a duration of \SI{2.5}{\s} (equivalent to microgravity duration).
The values given in Fig.~\ref{fig:GTBPgroundT} 
(and used throughout the remainder of the article)
are averages of $ \tau$ for $t \in [\SI{0.5}{\s}\text{ to }\SI{3}{\s}]$.
}

Even with this precaution,
the shear stress values are on average lower on ground those measured in microgravity (although not consistently),
\red{with $\langle \tau \rangle_{g_E} = \SI{5.67}{\pascal}$.
Stress fluctuations are significantly larger within individual tests ($c_v$ ranging from 23\% to 48\%),
but comparing the three repetitions, repeatability is higher than in microgravity ($c_v^{g_E}= 12\%$). 
}
The normal pressure under Earth gravity (Fig.~\ref{fig:GTBPgroundT}, lower panels) remains either roughly constant or decreases (up to $\SI{-40}{\pascal}$), indicating progressive densification of the packing under shear in Earth gravity.

\begin{figure}[htb!]
\includegraphics{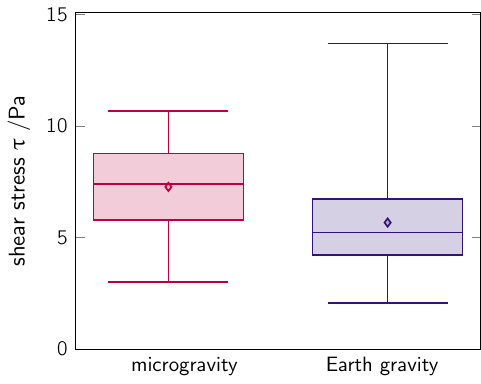}
	\caption{\label{fig:SS80-packing-torque-statistics}
\textbf{Statistics over
multiple shear tests conducted comparing microgravity to terrestrial gravity, with \SI{80}{\micro\meter} diameter polystyrene powder.} 
\red{
Boxplots use the data shown in Figs.\ \ref{fig:GTBPT} and \ref{fig:GTBPgroundT} (statistics are computed from all datasets combined, selected to be in microgravity and for a similar duration in Earth gravity -- see text for details).
Averages $\langle \tau \rangle_{g_\mu}$ and $\langle \tau \rangle_{g_E}$ are shown as diamond mark on each boxplot.
}
}
\end{figure}

Figure~\ref{fig:SS80-packing-torque-statistics} shows the statistics of shear stress
in both microgravity and Earth gravity, using boxplots
\red{
(where statistics are computed from the entire dataset).
It confirms that we measure generally higher shear stress in microgravity than in Earth gravity,
while stress fluctuations are higher when terrestrial gravity is present.
To assess the statistical significance of this difference, 
we compare the two samples using a $t$-test
(specifically, Welch's $t$-test, avoiding the equal variances assumption),
where the null hypothesis is that the two distributions have equal means within statistical variability, 
and the alternative hypothesis is that the shear stress is higher in microgravity,
$\langle \tau \rangle_{g_\mu } > \langle \tau \rangle_{g_E}$.
We find the difference between the datasets to be statistically significant 
at a 5\% significance level
($t(5.89) = 2.04$, corresponding to a $p\text{-value}=.0441$),
although the result is only just significant.
}


The second set of shear experiments uses larger particles ($d = \SI{1}{\mm}$),
and is conducted in the drop tower with catapult (Figure~\ref{fig:dtbtorque}(a)). 
In this configuration, shear is initiated only after the onset of microgravity. 
The average shear stress in microgravity reaches a notably high value of \SI{42.15}{\pascal}.
After the initial increase, the standard deviation remains around \SI{1}{\pascal}, 
similar to the ones found in previous microgravity experiments (Fig.~\ref{fig:GTBPT}).

Three independent repetitions are reproduced in terrestrial gravity (Figure~\ref{fig:dtbtorque}(b--d)). 
We find the average shear stress recorded on-ground to be lower than that in
microgravity, with $\langle \tau \rangle_{g_\mu } \approx 4 \langle \tau \rangle_{g_E}$.
\red{This increase in measured stress in microgravity 
follows the same trend as previously observed, but is quantitatively
much more pronounced.}
It confirms that granular materials subjected to shear in microgravity can exhibit markedly higher resistance to flow than under Earth gravity,
\red{and that we measure a significantly higher difference
for large, relatively lower-cohesion particles,
compared to particles small enough to already be considered cohesive under Earth gravity.
}

\section{Discussion}

\subsection{Jamming transition density}

Our compression experiment show that jamming transition appears at lower density in microgravity than on-ground: 
$\varphi_\text{J}(g_\mu) < \varphi_\text{J}(g_E) $, in agreement with previous work \cite{DAngelo2022} conducted on parabolic flights.
In these previous experiments, jamming was induced by a piston rising in a granular packing; 
we reproduce this result in compression, using larger particles (diameter \SI{1}{\mm} against \SI{80}{\micro\meter} in Ref.\ \cite{DAngelo2022}), and reinitializing the granular packing by fluidization between each test.
Here, we find a deviation of approximately 3\% between $\varphi_\text{J}(g_\mu) $ and $\varphi_\text{J}(g_E) $
(against 2\% in Ref.\ \cite{DAngelo2022}).
These deviations are small but statistically significant -- as confirmed below -- and hold practical relevance in the context of granular material processing, where slight changes in packing can markedly influence granular behavior.

Given the limited number of experimental data points available under microgravity, each measurement is particularly valuable. 
To evaluate the relation between $\varphi_\text{J}$ and $g$, 
we perform an unpaired $t$-test comparing the microgravity and terrestrial datasets
(all $\varphi_\text{J}$ values used are reported in Figure~\ref{fig:GTBPF}).
We find $t(11)=14.17$, corresponding to $p < .00001$;
this indicates that the difference is extremely statistically significant, supporting the conclusion that the packing fraction at jamming, $\varphi_\text{J}$, depends on gravitational acceleration.

\subsection{Rheology}

\begin{figure}
\centering
\includegraphics{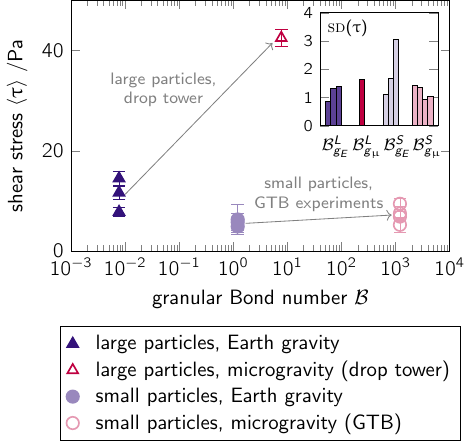}
\caption{\label{fig:stress-vs-bond}
\textbf{Summary of all rheology experiments.} 
Average shear stress, $\langle \tau \rangle$, as a function of the Bond number, $\mathcal{B}$, for all tested materials (large particles, \enquote{L}, and small particles, \enquote{S}) and gravitational conditions ($g_E$ and $g_\mu$).
Inset: bar plot of standard deviation of $\tau$, for each repetition of the experiment.
}
\end{figure}

Let us now consider the behavior of the materials under shear.
All experiments shown previously are summarized in Figure~\ref{fig:stress-vs-bond}, where the average shear stress, 
$\langle \tau \rangle$, is plotted as a function of the Bond number, $\mathcal{B}$, for all materials and gravity levels tested.
Across both materials and microgravity platforms, we consistently observe that a higher shear stress is required to deform the granular material under microgravity compared to Earth gravity.

We rationalize these findings using results from
\citeauthor{Murdoch2013}~\cite{Murdoch2013}, who showed that in microgravity, Taylor-Couette shear flow does not exhibit the secondary (radial) flow field typically present under gravity. 
In other words, under microgravity, particle rearrangement is restricted to the primary shear direction,
with no aid from gravity to facilitate reorganization.

On Earth, gravity acts as a supplementary load, normal to the shear direction. 
This provides additional confinement, but not only: it also introduces motion perpendicular to the shear direction, 
which can help particles reorganize. 
The absence of this mechanism in microgravity may therefore lead to enhanced resistance to shear, 
reflected in the increased stress measurements.

Besides, as discussed in Sec.~\ref{sec:methods}.\ref{sec:materials},
in microgravity, the absence of significant gravitational loading fundamentally alters the balance of forces acting on each particle.
On Earth, the weight of individual particles contributes to the overall confining pressure, effectively overcoming weak interparticle cohesive forces
in most dry granular systems \cite{Pouliquen2025}.
However, in microgravity, the gravitational load is minimized or eliminated, allowing cohesive forces to dominate particle interactions,
as highlighted by the shift in granular Bond number between $g_E$ and $g_\mu $ (\textit{cf.}\ Fig.\,\ref{fig:cohesionvasize}).
As a result, the granular assembly behaves more like a weakly bound solid than a loosely packed fluid \cite{Love2014,Gaida2025}. 
This shift in force balance partially explains the consistently higher shear stresses observed in microgravity experiments: 
the system behaves more like a cohesive medium, requiring greater stress to initiate and sustain shear. 
Furthermore, the predominance of cohesion over weight reduces particle sliding and rearrangement.

Now comparing the two particle species, we find a fourfold increase in $\langle \tau \rangle$ for large particles in the drop tower microgravity experiments, but only a \red{22\%} increase for small particles in the \gls{gtb} microgravity experiments,
both compared to $\langle \tau \rangle_{g_E}$.

We propose two explanations. The first one is technical, 
arising from the use of different low gravity platforms. 
In the \gls{gtb}, active \enquote{drops} generate stronger vibrations and thus higher $g$-jitters, 
yielding microgravity of lower quality than in the passive higher drop tower, by roughly two orders of magnitude \cite{Gierse2022}. 
These vibrations facilitate particle reorganization, reducing resistance to flow and thereby decreasing the measured shear stress.

The second explanation concerns the nature of the granular materials themselves. 
Comparing the two spices' Bond numbers to quantify the balance of cohesive to inertial forces, 
for small particles, cohesion is already relevant under Earth gravity ($\mathcal{B}^S_{g_E} \approx 1$).
The material becomes \emph{even more} cohesive in reduced gravity ($\mathcal{B}^S_{g_\mu} \approx 10^3$).
For large particles, by contrast, $\mathcal{B}^L_{g_E} \approx 10^{-2} \ll 1$, so inertial forces predominate on Earth. 
In microgravity, however, $\mathcal{B}^L_{g_\mu} > 1$: the material effectively \emph{becomes} cohesive -- which may explain the much larger increase in $\langle \tau \rangle$.

\subsection{Stress fluctuations}

After comparing variations across repetitions of the same experiment 
(\textit{cf.}\ Fig.\,\ref{fig:SS80-packing-torque-statistics}), 
we now examine stress fluctuations within each dataset, for all experiments.
The inset of Fig.\,\ref{fig:stress-vs-bond} shows a bar plot of the standard deviation for each shear test, $\textsc{sd}(\tau)$.

For the large particles, $\textsc{sd}(\tau)$ is nearly identical at both gravity levels,
albeit slightly higher in $g_\mu $,
but having only one microgravity repetition prevents firm conclusions.

For the small particles, $\textsc{sd}(\tau)$
is \red{generally} larger in Earth gravity than in microgravity.
Stress fluctuations can also be characterized by the peak frequency, $ \f_p $. 
Consistent with $\textsc{sd}(\tau)$, we find
a significantly higher frequency of fluctuations for measurements conducted on ground:
$\langle \f_p \rangle_{g_E } = \SI{2.08}{\hertz}$, 
against $\langle \f_p \rangle_{g_\mu } = \SI{1.52}{\hertz}$ for our microgravity experiments.

Our observations align with findings by \citeauthor{Featherstone2021}~\cite{Featherstone2021}, who report more frequent stick-slip events at higher gravity.
In contrast, granular media in microgravity were described as more \emph{fluid-like}, with smoother and more continuous deformation of the granular fabric. 
This reduced variability in our microgravity data supports such interpretation: without the gravity-induced normal stress, 
the contact network might be less prone to intermittent locking and sudden release (i.e., stick-slip).

At the microscopic scale, 
the particles can be described as cohesively bound with limited friction, 
allowing them to remain in contact while still sliding and/or rolling on each other. 
As in our experiments, contacts are already established at the onset of shear, the flow remains relatively continuous. 
In Earth gravity, however, the additional load direction can promote denser regions opposing shear, 
amplifying local heterogeneities and stick-slip behavior, hence 
stronger fluctuations.

Note, however, that in this study we used only relatively smooth, spherical particles; increasing surface friction, roughness, and/or introducing angular or interlocking shapes might reveal additional features of granular flow behavior in microgravity.

\subsection{Granular bed expansion}

\begin{figure}
\centering
\includegraphics{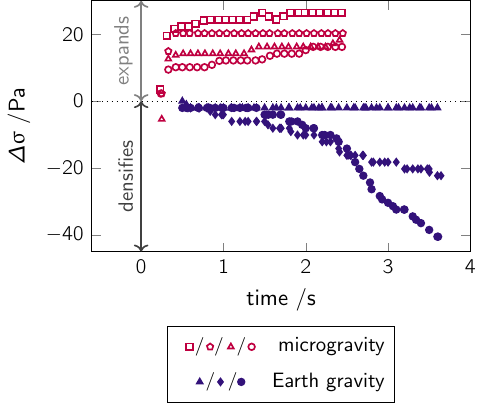}
\caption{\label{fig:TimeDependentLoadsEvolution}
\textbf{Evolution of relative normal pressure in Earth- and microgravity.}
The normal pressure change, $\Delta\sigma$, applied by the granular media to the piston,
is plotted as a function of time for all experiments with small particles ($d=\SI{80}{\micro\meter}$).
Microgravity experiments were conducted in the \gls{gtb}.
Different symbols represent the different repetitions of the experiment
(data shown in Figs.\ \ref{fig:GTBPT} and \ref{fig:GTBPgroundT}).
}
\end{figure}

A closer look at the evolution of the normal pressure during shear 
(Fig.~\ref{fig:TimeDependentLoadsEvolution})
also shows that in microgravity, the granular packing expands over time, whereas on Earth, it densifies. 
This contrast is again consistent with the absence of gravity-induced settling in microgravity: 
once particles are agitated by shear, there is no preferential direction to fall into interstitial voids and compact the granular packing. 
On Earth, however, gravity forces particles to rearrange into denser configurations during shear (as well as during compression), promoting densification.
This markedly different response suggests that granular rheology is gravity-dependent; for instance, applying a rheological law such as $\mu(I)$ \cite{Jop2006}, would yield very different effective friction coefficients, even for the same material, in different $g$.

\subsection{Summary: granular flows \textit{vs.}\ gravity}

To summarize,
we propose the following hypotheses to rationalize our results:
\begin{itemize}
\item \textbf{Minimization of potential energy:} 
In a gravitational field, particles tend to move to lower positions to minimize their potential energy, 
provided the packing geometry allows for it. This leads to a net densification of the granular medium, 
as gravity drives downward reorganization of particles.

\item \textbf{Gravity as a supplementary load and loading direction:} 
Gravity provides a secondary force field, adding a normal stress component to the 
applied stress tensor \cite{Murdoch2013, DAngelo2022}.
This might promote plastic rearrangement by helping particles overcome friction and escape local jamming. 
On Earth, such gravity-induced secondary force field assists reorganization; 
its absence in microgravity might results in increased resistance to shear.

\item \textbf{Reduced variability and stick-slip dynamics in microgravity:} 
Gravity enhances local heterogeneities and promotes intermittent stick-slip behavior \cite{Featherstone2021}. In contrast, granular media in microgravity behave more fluid-like, with smoother, more continuous deformation, 
possibly due to reduced frictional locking in the absence of gravity-induced normal loading.

\item \textbf{Enhanced role of cohesive forces in microgravity:}
Interparticle cohesive forces (e.g., van der Waals, electrostatic, capillary) are typically much weaker than gravity;
in microgravity, they are no longer dominated by particles' weight and thus become predominant~\cite{Love2014, Gaida2025}. 
These cohesive interactions resist particle separation, which might result in granular materials requiring higher stress to shear.

\end{itemize}

\section{Conclusion}

In this study, we investigated the influence of gravity on granular density at jamming
and shear response,
through a series of experiments conducted in microgravity and terrestrial gravity conditions.
We used an in-house developed setup combining compression and shear (Taylor-Couette shear geometry), 
on two microgravity platforms: the \gls{gtb} (smaller active drop tower) and the drop tower with catapult at the \gls{zarm} in Bremen, Germany.
For materials, spherical, monodispersed polystyrene particles of two sizes were employed ($d=\SI{80}{\micro\meter}$ and $d=\SI{1}{\mm}$).

Across all tests, we consistently found
that the packing fraction at jamming, $\varphi_\text{J}$, depends on the gravitational acceleration,
with granular packing jamming at lower packing density in microgravity than in Earth gravity:
$\langle \varphi_\text{J} \rangle_{g_\mu}= 0.587$ and $\langle \varphi_\text{J} \rangle_{g_E} = 0.606$, respectively.
This result is statistically significant, as evidenced by a
an unpaired T-test 
\red{($p$-value below $.00001$).}

Regarding granular rheology in microgravity,
we find that higher shear stress is required 
in microgravity than in terrestrial gravity,
with $\langle \tau \rangle_{g_\mu} = \SI{7.27}{\pascal}$ in \gls{gtb} experiments and \red{$\langle \tau \rangle_{g_E} = \SI{5.67}{\pascal}$} in terrestrial gravity.
\red{We find the difference to be statistically significant
(although weaky), with a $p$-value of $.0441$.}
We also find reduced variability in microgravity measurements, indicating smoother, more continuous flow.

We propose different hypothesis of gravity-dependent mechanisms to explain our results.
On Earth, gravity not only imposes an additional load normal to the shear plane, 
but also enables particle rearrangement through a secondary force field.
In contrast, in microgravity, its absence might lead to reduced plastic reorganization and increased resistance to shear. Furthermore, cohesive forces -- sometimes negligible compared to particles' weight -- become predominant in microgravity, further impeding particle rearrangement and enhancing shear resistance.
\red{This effect appears to be strongest when the change in gravitational environment shifts the granular Bond number from below to above 1,
but the limited dataset available warrants further investigation.}

Together, our results reveal how gravity affects not only the magnitude of stresses required to shear granular media, 
but also the underlying particle dynamics. 
These insights are important for predicting and controlling granular behavior in low gravity environments, with direct implications for planetary exploration, \gls{isru}, and spacecraft operations involving regolith or granular flows.

Future work should aim to build on these findings through more systematic and comprehensive experiments (as proposed for example by \citeauthor{Duffey2025}~\cite{Duffey2025}), 
including longer-duration experiments, and using a broader variety of granular materials;
this includes varying particle shape, size distribution, surface roughness, and cohesive properties, to better isolate and quantify the influence of each factor. 
Low- and microgravity experiments provide an opportunity to expose microscopic interactions typically masked by Earth's dominant gravitational force.
Ultimately, such efforts aim to develop a general granular rheology model that remains valid under reduced gravity.

\nolinenumbers
\printbibliography

\section*{Acknowledgements}
We would like to extend gratitude to Achim Sack and Walter Pucheanu for their help in creating the experimental setup,
as well as the \gls{zarm} personnel, in particular Dieter Bischoff.

This work was supported by the German Aerospace Center (DLR) Space Administration 
with funds provided by the German Federal Ministry 
for Economic Affairs and Climate Action (BMWK)
under grant number 50WM2342A
and through DLR-granted access to the ZARM drop tower facilities and engineering support under the project name 
\enquote{J \textit{versus} $g$}.

O.\ D'A.\ acknowledges financial support from 
the Emerging Talents Initiative (ETI) program 2023 by the
Friedrich-Alexander-Universit\"{a}t (FAU) Erlangen-N\"{u}rnberg,
and from the French National Centre
for Space Studies (CNES) under the CNES fellowship 24-357
and APR ID~10678 (2025).

\section*{Data Availability Statement}
Data supporting this study are openly available from 
the Zenodo repository 17250245
(\href{https://doi.org/10.5281/zenodo.17250245}{DOI: 10.5281/zenodo.17250245}) \cite{ZenodoRep}.


\end{document}